\documentclass[]{tLCT2e}
\usepackage{graphicx}
\usepackage{color}

\begin{document}
\doi{10.1080/0267829YYxxxxxxxx}
\issn{1366-5855}  \issnp{0267-8292}
\jvol{00} \jnum{00} \jyear{2012} \jmonth{July}

\markboth{Taylor \& Francis and I.T. Consultant}{Liquid Crystals}

\articletype{Invited Article}

\title{Micro-Bullet Assembly: Interactions of Oriented Dipoles in Confined Nematic Liquid Crystal}


\author{Mohamed Amine Gharbi$^{\rm a}$$^{\rm b}$$^{\rm c}$, Marcello Cavallaro Jr.$^{\rm b}$, Gaoxiang Wu$^{\rm c}$, Daniel A. Beller$^{\rm a}$, Randall D. Kamien$^{\rm a}$, Shu Yang$^{\rm b}$$^{\rm c}$ and Kathleen J. Stebe$^{\rm b}$$^{\ast}$\thanks{$^\ast$Corresponding author: kstebe@seas.upenn.edu. 
\vspace{6pt}}\\{\vbox{\vspace{12pt} $^{\rm a}${\em{Department of Physics and Astronomy, University of Pennsylvania, Philadelphia, PA 19104-6396, USA}}; $^{\rm b}${\em{Department of Chemical and Biomolecular Engineering, 220 S. 33rd Street, Towne Bldg. Rm. 364, Philadelphia, PA 19104, USA}}; $^{\rm c}${\em{Department of Materials Science and Engineering, University of
Pennsylvania, 3231 Walnut Street, Philadelphia, PA 19104, USA.}}\\\vspace{6pt}\received{v1.0 released January 2008}}} }


\maketitle

\begin{abstract}

Microbullet particles, cylinders with one blunt and one hemispherical end, offer a novel platform to study the effects of anisotropy and curvature on colloidal assembly in complex fluids. Here, we disperse microbullets in 4-cyano-4'-pentylbiphenyl (5CB) nematic liquid crystal (NLC) cells and form oriented elastic dipoles with a nematic point defect located near the curved end. This feature allows us to study particle interactions as a function of dipole alignment. By careful control of the surface anchoring at the particle surface and the confining boundaries, we study the interactions and assembly of microbullets under various conditions. When microbullets with homeotropic surface anchoring are dispersed in a planar cell, parallel dipoles form linear chains parallel to the director, similar to those observed with spherical particles in a planar cell, while antiparallel dipoles orient side-to-side. In a homeotropic cell, however, particles rotate to orient their long axis parallel to the director. When so aligned, parallel dipoles repel and form 2D ordered assemblies with hexagonal symmetry that ripen over time owing to attraction between antiparallel neighbors. Further, we show that the director orientation inside the cell can be altered by application of an electrical field, allowing us to flip microbullets to orient parallel to the director, an effect driven by an elastic torque. Finally, we detail the mechanisms that control the formation of 1D chains and hexagonal lattices with respect to the elasticity of the NLC.

 \bigskip

\bigskip

\end{abstract}

\section{Introduction}

The self-assembly of colloidal particles has stimulated great scientific interest, as it serves as a model system to answer fundamental questions in crystallization \cite{Pusey1986}, phase transitions  \cite{Kooij2000}, and topology \cite{Terentjev1995,Poulin1997, Lubensky1998, Abott2000, Stark2001,Stark2002}. When colloidal particles are dispersed in complex fluids - such as liquid crystals (LCs) that are inherently anisotropic, with physical properties dependent on molecular alignment - advanced materials with novel mechanical, optical or electronic properties are expected \cite{Mende,Hwang2005, Sergeyev,Hung2009}. Of specific interest are the rod-like nematic liquid crystal (NLC) molecules that tend to align their major axes along one direction known as the director. 

The anisotropy of NLCs has been exploited to direct colloidal assembly \cite{Poulin1998, Loudet2000, Musevic2006, Nych2007, Skarabot2008, Musevic2008a, Ognysta2009}. The dispersion of colloidal particles in the NLC can strongly disturb the local director field around the particles, leading to an elastic response that is markedly different from that in simple fluids, where typically only isotropic forces come into play \cite{Barrat2003}. The degree of deformation of the LC's local director depends on surface anchoring conditions \cite{Terentjev1995,Poulin1998,PoulinA99}, particle size  \cite{Abott2000}, shape \cite{Lapointe2009, Loudet2009}, and confinement \cite{Skarabot2008, Vilfan2009}. Collectively, these effects lead to spatial discontinuities of the director field, known as topological defects, which are regions where the NLC molecules are locally disordered. The defects form in order to satisfy LC anchoring conditions on the particle and director orientations imposed by confinement, and can be approximated in the far field as multipole charges, by analogy with electrostatics \cite{Lubensky1998}. For example, when spherical colloids with homeotropic anchoring are present in oriented NLCs, defects form with either dipolar or quadrupolar symmetry. In a dipolar configuration, the colloid is accompanied by a point defect called a hyperbolic hedgehog, whereas for a quadrupolar configuration, a disclination loop known as a Saturn ring encircles the particle. In dense particle systems, a competition between elasticity, topological defects, and anchoring energy gives rise to long-range interactions responsible for the formation of ordered structures ranging from 1D chains to 2D crystals \cite{Poulin1997, Poulin1998, Loudet2000, Musevic2006, Kotar2006, Ognysta2008, Ognysta2011}.

While spherical colloids in NLCs have been extensively studied, recent advances in microfabrication techniques make it possible to study complex shaped particles in interaction with LCs \cite{Lapointe2009, Loudet2009, Smalyucka2011, Smalyuckb2011, Smalyuck2012, Smalyuck2012b, Smalyuck2012c, Musevic2008}. It has been suggested that such interactions depend strongly on particle shape. For example, polygonal platelet particles with planar surface anchoring form different assemblies depending on the number of particle sides when dispersed in NLC \citep{Lapointe2009}. In addition, particle orientation can be manipulated by an external field that distorts the director field \cite{Lapointe2004, Lapointe2005, Lapointe2008}, providing a new means for tuning particle interactions leading to new assemblies and applications of colloids in LCs.

In this work, we fabricated rod-like ``bullet"-shaped microparticles rounded at one end and flat at the other, and studied their behavior in 4-cyano-4'-pentylbiphenyl (5CB) NLC. We carefully controlled the anchoring conditions of the NLC at the particle surfaces and the confining boundaries. Microbullets with homeotropic anchoring dispersed in a uniform nematic director field created defects in order to satisfy global topological constraints and behaved as elastic dipoles with a particular orientation associated with the particle shape. This allowed the study of particle assembly as a function of dipole orientation in liquid crystal cells with either planar or homeotropic surface anchoring. We probed pair interactions in sparse systems and elucidated collective mechanisms that drive assembly into 1D and 2D structures. The prominent role of the  director field in determining particle orientation  was shown  by re-orienting the director using an applied electric field.

\section{Materials and Methods}
\subsection{Fabrication of Microbullets}

A polydimethylsiloxane (PDMS) mold was prepared by replica molding from a Si master consisting of a micropost array (radius of $R$ = 1 $\mu$m and height, $L$ = 10 $\mu$m) with rounded tops following the literature \cite{Zhang2006}. A mixture of 40 \% wt. silica nanoparticles of diameter 100 nm (ORGANOSILICASOL$^{TM}$, Nissan chemicals) dispersed in ethoxylated trimethylolpropane triacrylate (ETPTA, Sigma-Aldrich) was prepared according to the literature \cite{Yang2004} and infiltrated into the PDMS mold by capillary force lithography (CFL) \cite{Zhang2010}, followed by UV curing at 365 nm with exposure dosage of 5000 mJ/cm$^2$ to create ETPTA/silica nanoparticle micropost arrays.

\begin{figure}
\begin{center}
\includegraphics[width=0.8\textwidth]{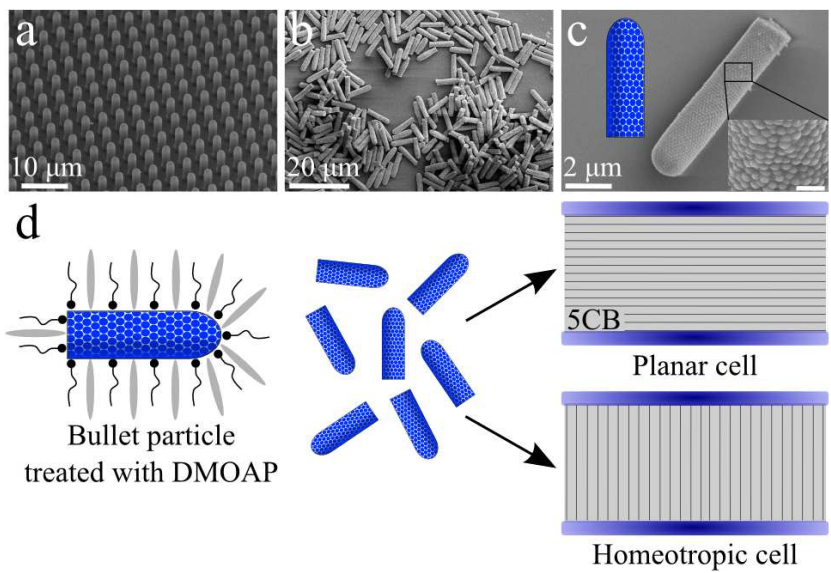}
\caption{SEM images (a) of a micropillar array on a silicon substrate; (b) of individual microbullets after
release from the silicon wafer; (c) showing the bullet-shape of a colloid covered by silica nanoparticles ($L = 10$ $\mu$m in length and $R$ = 1 $\mu$m in radius). Inset: SEM image showing the surface roughness of the microbullet (scale bar is 400 nm). (d) Schematic of a microbullet particle with homeotropic anchoring to be dispersed in planar and homeotropic NLC cells.}
\label{Fig1} 
\end{center}
\end{figure}

To avoid the inhibition of the free radical polymerization of ETPTA due to the oxygen dissolved in the PDMS mold, the entire CFL process was performed under an inert atmosphere, argon gas. After CFL, the micropost arrays were obtained by carefully peeling off the PDMS mold. During the CFL process, silica nanoparticles were arranged into densely packed structures at the solid-liquid interface covering the outer surface of the microbullet. The presence of silica nanoparticles allowed surface functionalization to dictate surface anchoring of NLCs. To expose the nanoparticles, the surface polymers were partially removed via oxygen reactive ion etching (RIE), thereby creating surface roughness depending on the concentration of the silica nanoparticles and the etching depth. In general, the sample fabricated from the dispersion with higher silica nanoparticle concentration has higher surface roughness due to the denser packing of the silica inside the micro-post. Higher surface roughness is also expected for larger etching depth because more silica nanoparticles can be exposed the surface of the micro-posts; a detailed study of nanoparticle assembly within the microposts will be presented elsewhere \cite{Wu2012}. To impose homeotropic anchoring at the particle surface, we treated the microposts with a 3\% wt. solution of N,N-dimethyl-N-octadecyl-3-aminopropyltrimethoxysilyl chloride (DMOAP, Sigma) in a mixture of  90\% wt. ethanol and 10\% wt. water as described in \cite{Skarabot2008,Gharbi2011}. Thereafter, the microposts were liberated from the substrate by scraping using a razor blade to generate microbullet particles. An additional surface functionalization step was used to ensure the flat sides of the microbullets were also coated with DMOAP. Finally, we dried particles in a vacuum oven at $T$ = 110 $^\circ$C for an overnight before use in experiment.

\subsection{Preparation of Liquid Crystal (LC) Cells}

4-pentyl-4'-cyanobiphenyl (5CB, Kingston Chemicals Limited), which forms a nematic phase at room temperature, was used in the experiments. We fabricated LC cells with either oriented planar or homeotropic anchoring. To ensure oriented planar anchoring, clean glass slides were treated with a 1\% wt. solution of polyvinyl alcohol (PVA, Sigma-Aldrich) in a mixture of 95\% wt. water and 5\% wt. ethanol, and rubbed along one direction after heating at $T$ = 110 $^\circ$ C for one hour. Rubbing directions at the top and bottom substrates were antiparallel to prepare a uniform director. To impose homeotropic anchoring, clean glass slides were treated with 0.1\% wt. DMOAP solution in a mixture of 90\% wt. ethanol and 10\% wt. water and subsequently heated at 110 $^\circ$ C overnight. The thickness of the LC cells $h$ was 25 $\mu$m defined by Mylar spacers. For application of an electric field, glass slides coated with indium tin oxide (ITO, 70-100 $\Omega$/sq, Sigma-Aldrich) were used with oriented planar anchoring imposed as described above. An NLC/microbullet suspension (1\% - 3\% wt. of microbullets in 5CB) is then introduced into the LC cells via capillarity.

\subsection{Characterization}

Microbullets were characterized using scanning electron microscopy (SEM). The SEM images, as shown in Figure \ref{Fig1}, were taken by a FEI Strata DB235 Focused Ion Beam in high vacuum mode with acceleration voltage of 5kV. The particles were named because of their shape resembling bullets, with a cylindrical body having one rounded end and one relatively flat end. The microbullets in the NLC system were observed under an upright optical microscope (Zeiss AxioImager M1m) in transmission mode. The optical microscope was equipped with a heating stage (Bioscience Tools, temperature regulated at 0.1 $^\circ$C) and a set of crossed polarizers. Images were recorded with a high-resolution camera (Zeiss AxioCam HRc) and high-speed camera (Zeiss AxioCam HSm). We used a function generator (DS340, Stanford Research Systems) to apply sinusoidal voltage $U$ (0 $\leq$ $U$  $\leq$ 10 V) at 1 kHz across the ITO glass of the cell. Finally, particle trajectories over time were tracked using ImageJ.

\section{Results and Discussion}

\begin{figure}
\begin{center}
\includegraphics[width=0.8\textwidth]{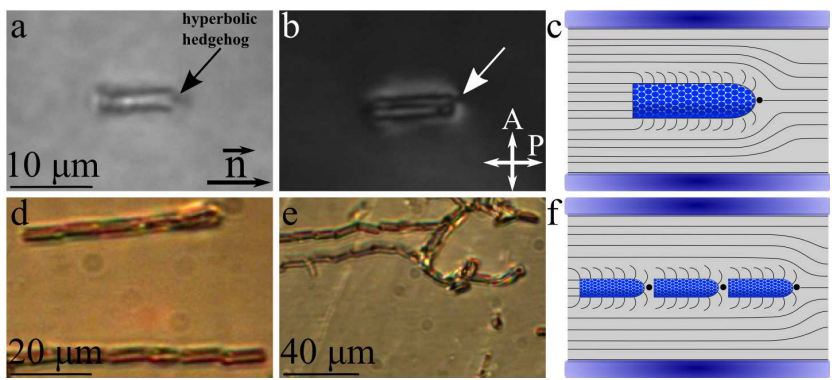}
\caption{Dispersion and assembly of microbullet particles in an NLC planar cell. (a-b) Optical images of an isolated microbullet dispersed in a planar cell. The presence of a point defect close to the spherical cap of the particle observed in (a) bright field and (b) between crossed polarizers (b), respectively. (c) Schematic representation of the dipolar structure of the director around an isolated microbullet. (d) Optical image of 1D chains of microbullets along the easy axis of the cell. (e) Optical image of complex structures of microbullet chains. (f) Schematic representation of a possible nematic texture around a chain of microbullets.}
\label{Fig2} 
\end{center}
\end{figure}

Isolated microbullets in planar LC cells assumed a preferred orientation with their long axis parallel to the far-field director set by the anchoring conditions at the boundaries. Under crossed polarizers, a dipolar defect structure was revealed, with a point defect always present near the rounded end of the particle (Figure \ref{Fig2} (a)-(b)), just as hyperbolic hedgehog defects form near homeotropic spherical beads in planar cells \cite{Poulin1998}. At low density, there was a high population of interacting parallel dipoles that assembled into 1D chains parallel to the director (see Figure \ref{Fig2} (d)). As the chains grew in size, end-to-end chaining was hindered and large disordered structures began to form due to the strong interactions between chains (Figure \ref{Fig2} (e)). These interactions can be attributed to the inherent elasticity in the nematic phase. When the samples were heated above the nematic-isotropic transition point of 5CB ($T_{NI}$ = 34 $^\circ$C), the defect structures disappeared and pair interactions were lost. If samples were subsequently cooled back to the nematic phase, the aforementioned behavior was restored. Schematic representations of the overall director field for an isolated microbullet and a chain of microbullets are provided as insets in Figure \ref{Fig2} (c) and Figure \ref{Fig2} (f), respectively, to illustrate the behaviors of microbullets in planar cells.

\begin{figure}
\begin{center}
\includegraphics[width=0.8\textwidth]{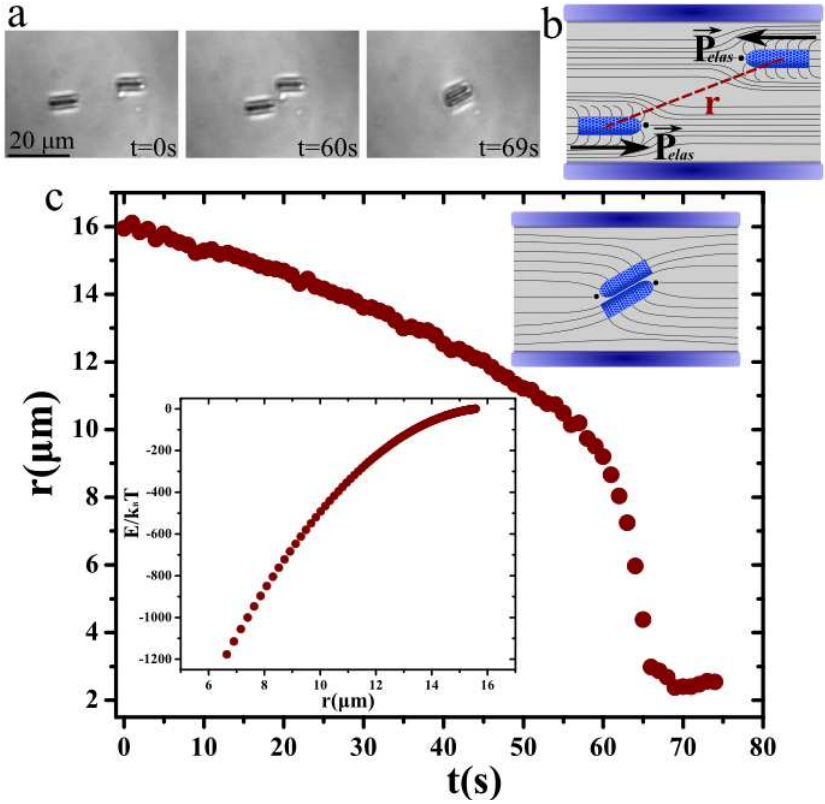}
\caption{Interactions between a pair of antiparallel microbullets in a planar NLC cell. (a) Sequential optical images capturing interactions of antiparallel microbullets. (b) Schematic representation of the nematic director field for two particles in proximity. The centers of mass of the microbullets are separated by a distance $r$. (c) Time dependence of the separation distance between two isolated microbullets and (inset) the corresponding pair potential derived from the trajectory in experiment. Inset: the possible LC texture around microbullets when coming into contact. }
\label{Fig3} 
\end{center}
\end{figure}

Antiparallel dipoles also attracted in the far field with dynamics similar to the parallel dipoles. However, when they were in close proximity (i.e. when the distance between the centers of mass of the microbullets, $r$, was approximately 3 $\mu$m), they slide into a side-to-side arrangement rather than forming chains (see Figure \ref{Fig3}).

Much of this behavior can be understood by analogy with spherical particles with strong homeotropic anchoring dispersed in planar LC cells. The homeotropic microbullet is topologically identical to the homeotropic sphere, as can be seen in Figure \ref{Fig2} (c), so the microbullet acts as a radial hedgehog defect which can be assigned, for instance, a topological charge of $s$ = +1. The accompanying point defect has a hyperbolic hedgehog geometry with opposite topological charge $s$ = -1 \cite{Kamien2012}.

\begin{figure}
\begin{center}
\includegraphics[width=1\textwidth]{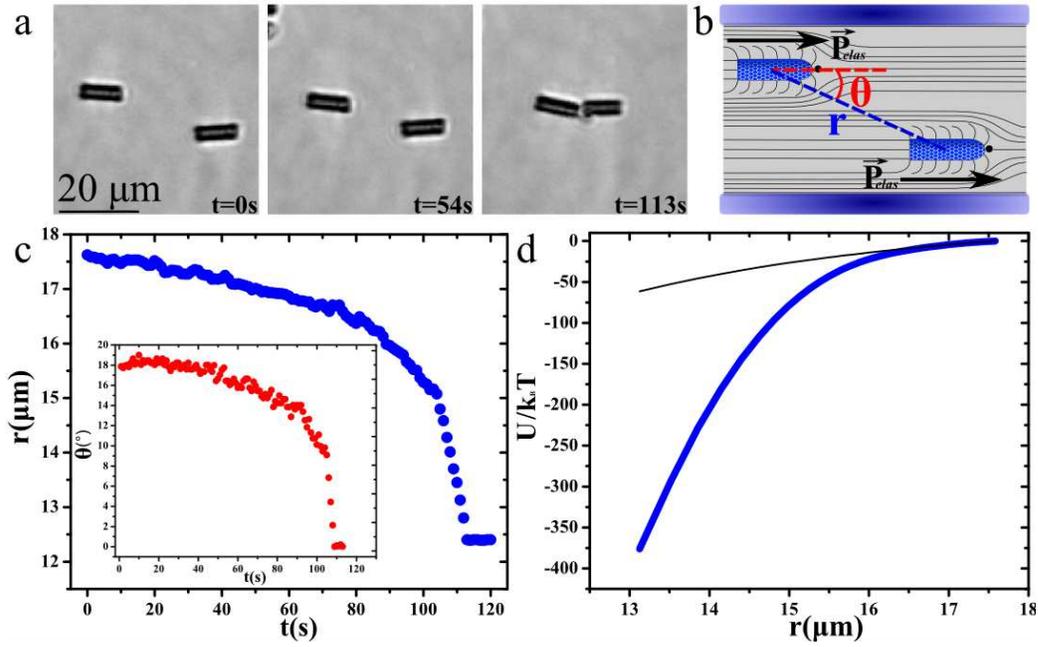}
\caption{Interaction between a pair of microbullets in a planar NLC cell. (a) Sequential images capturing interactions of parallel microbullets. (b) Schematic representation of the nematic director field for two particles in proximity. The centers of mass of the microbullets were separated by a distance $r$ and the line connecting their centers of mass make an angle $\theta$ with the easy axis of the NLC cell. (c) Time dependence of the separation distance between two isolated bullets and their separation angle (inset). (d) Corresponding pair potential derived from the trajectory in experiment and arbitrarily fixed at 20 $\mu$m (thick blue line) and its comparison with an elastic dipolar model at large distances (thin black line).}
\label{Fig4} 
\end{center}
\end{figure}

To analyze particle pair interactions, we used video microscopy to record trajectories and track separation distances over time. Typical experiments were performed at room temperature $T$ = 21 $^\circ$C. A trajectory of a pair of particles for initial separation distances of roughly $r \approx$ 20 $\mu$m and an initial angle of $\theta$ $\approx$ 18$^\circ$ for parallel dipoles is shown in Figure \ref{Fig4}. When the Ericksen and Reynolds numbers are small, the motion of particles can be approximated as resulting entirely from the LC-mediated elastic interaction potential between microbullets \cite{Galerne2007}. In this work, the Ericksen number $Er$ and  Reynolds number $Re$ are:
\begin{equation} 
Er = \eta v L_{c} / K \approx 2\times 10^{-3} ; Re = \rho v L_{c} / \eta \approx 10^{-7}
\end{equation}
where $\eta= 14.21$ mPa s is the viscosity of 5CB, $v=0.13$ $\mu$m s$^{-1}$ is the particle velocity, $L_c$ is the characteristic length (the length of the microbullet), $K \approx$ $10^{-11}$ N is the average Frank elastic constant, and $\rho$ =1.05 g cm$^{-3}$ is the density of 5CB. The interaction force $f_p$ can then be obtained from Stokes's law, where $\gamma$ is the drag coefficient. We deduce $\gamma$ from the diffusion coefficient $D$ defined by $D=k_BT/\gamma$ as obtained from a mean square displacement analysis of an isolated microbullet. The measurement of $D$ in both directions, parallel $D_{\parallel}$ and perpendicular $D_{\perp}$ to the director, showed that the diffusion of microbullets was anisotropic. Bullets diffused faster along the director ($D_{\parallel} = 6.8 \times 10^{-4} \mu$m$^2$ s$^{-1}$) than in perpendicular directions ($D_{\perp} = 4.8 \times 10^{-4} \mu$m$^2$ s$^{-1}$) due to the alignment of the director field and the anisotropic shape of particles. For the purposes of estimating the pair potential, an  average of these two diffusion coefficients was used. A typical pair interaction landscape is given in Figure \ref{Fig4} (d).

In the far field, the microbullet trajectories can be described as two interacting elastic dipoles \cite{Lubensky1998}. In the one-constant approximation, the elastic free energy of the NLC/bullet system is given by:
\begin{equation} 
U_{el}=\frac{K}{2}\int {\rm d} ^3r(\nabla n)^2
\end{equation}
where $n$ is the unit nematic director. The coordinate system is chosen such that the $x$ axis aligns with the far-field director to give the lowest order $n \approx (1,n_y,n_z)$. The Euler-Lagrange equation arising from the minimization of the elastic free energy reduces to the Laplace equation for $n$:  
\begin{equation} 
\nabla ^2 n^i=0, i=y,z
\end{equation}
At a large distance $r$, the solution for Eq.2 is given by a multipole expansion in $n$:
\begin{equation} 
n^i=\frac{A^i}{r}+\frac{P_j^i r_j}{r^3}+\frac{Q^i_{jk} r_j r_k}{r^5}+...
\end{equation}
where $A$, $P$ and $Q$ are the elastic monopole, dipole and quadrupole moments, respectively, of the bullet-defect combination. Summation over $j$ and $k=x,y,z$ is implied. To leading order, and assuming rotational symmetry about the $x$-axis, this topological dipole configuration gives:
\begin{equation} 
n^i=\frac{Pr^i}{r^3}, i=y,z
\end{equation}
Dimensional analysis requires the magnitude $P$ of the elastic dipole moment to be proportional to a distance squared. Since the deformation of the director field around the particle should depend on the particle dimensions (the length $L$ and the radius $R$ of the microbullet), we can deduce that $P\sim LR$.  The associated long-range pair interaction is:
\begin{equation} 
U_{el} \propto K \frac{(LR)^2}{r^3}(1-3cos^2 \theta)
\end{equation}
where $r$ is the distance between the centers of mass of microbullets and $\theta$ is the angle between the line connecting the centers of particles and the easy axis (Figure \ref{Fig4} (b)). At a large distance (typically, $r \geq$  16.5 $\mu$m and $t \leq$ 65 s), we observed that the variation in $\theta$ was negligible, $\theta =(17.5 \pm 1.7) ^{\circ}$. For a constant $\theta$, the potential is reduced to:
 \begin{equation} 
U_{el} \propto -K \frac{(LR)^2}{r^3}
\end{equation}

The pair interaction energy extracted from the experimental data (bold line in Figure \ref{Fig4} (d)) agrees well with the prediction at a large separation distance, confirming that an elastic dipole interaction is responsible for the long-range attraction in this system. However, the model breaks down at short separation distances due to the significant variation of $\theta$ for separations below 16 $\mu$m and significant distortions in the local director field.

In our second system, we introduced microbullet particles into homeotropic cells ($h$ = 25 $\mu$m). Microbullets rotated upon dispersion in the cell and oriented their long axis perpendicular to the glass slides, along the director of the NLC. When the sample was heated beyond $T_{NI}$ particles became randomly oriented.

\begin{figure}
\begin{center}
\includegraphics[width=0.6\textwidth]{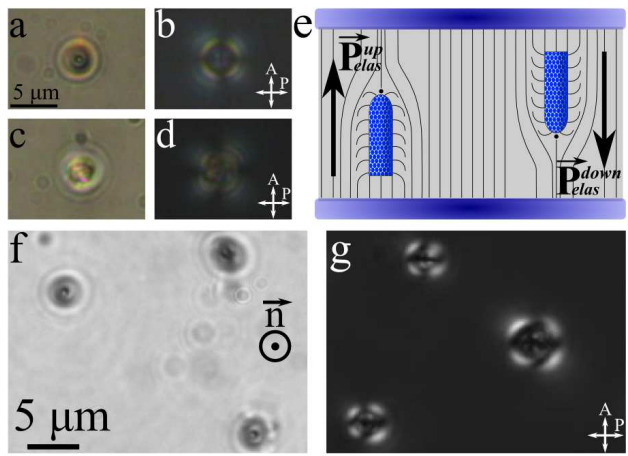}
\caption{Microbullets in a homeotropic NLC cell. (a) Bright field and (b) polarized optical microscopy images showing the hyperbolic hedgehog defect for an upward pointing elastic dipole. (c) Bright field and (d) polarized optical microscopy images showing the hyperbolic hedgehog defect for a downward pointing elastic dipole. (e) Schematic representation of the nematic director field around an upward and a downward pointing elastic dipole. (f) Bright field and (g) polarized optical microscopy images showing the presence of up and down elastic dipoles in the same NLC cell.}
\label{Fig5} 
\end{center}
\end{figure}

We examined the topological details of the director field around individual bullets using polarized light microscopy. The observed birefringence patterns and a schematic of the possible structure of the director field are provided in Figure \ref{Fig5}. These images indicate that the global director structure of the nematic was dipolar as in the case of planar cells, with the presence of a hyperbolic hedgehog defect localized close to the hemispherical cap of the bullet. We were also able to determine the orientation of the dipole (spherical cap pointing up or down), which has implications for the particle position and interactions, as discussed below. The positioning of the microbullet was dictated by the confinement of the cell walls as well as the imposed homeotropic surface anchoring. Between crossed polarizers, a Maltese cross pattern was observed at a different plane along the $z$-axis when the dipole was up (defect above microbullet) vs. when the dipole was down (Figure \ref{Fig5} (a) - (d)). The microbullet sat at a lower height in an upward-pointing dipole than in a downward-pointing dipole. The typical shift in height for our system was roughly $d$ $\approx$ 1.5 $\mu$m (Figure \ref{Fig5} (f) and Figure \ref{Fig5} (g)). From Figure \ref{Fig5} (e), it seems geometrically reasonable to expect that the flat edge of the microbullet would be situated closer to a flat homeotropic boundary than would the hyperbolic hedgehog defect. These results indicate that the asymmetry of the microbullet structure offers unique control over the height of the particle, which is not possible with spherical particles.	
	
\begin{figure}
\begin{center}
\includegraphics[width=0.6\textwidth]{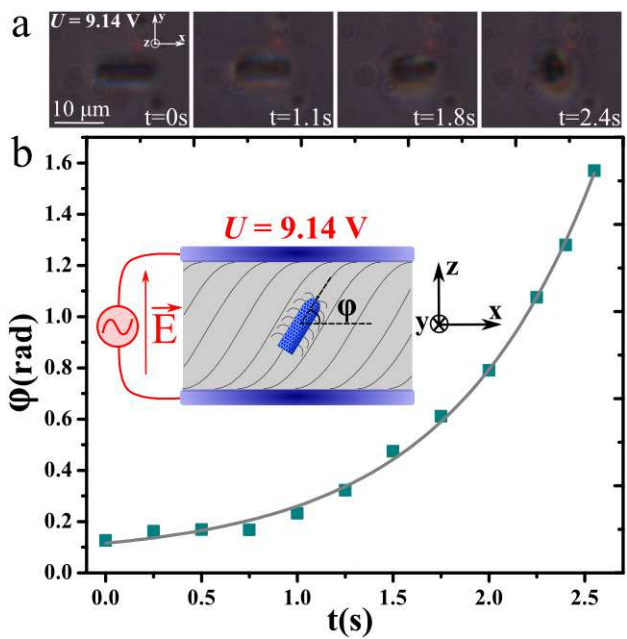}
\caption{Sequential optical images of rotation of a microbullet in response to an applied electrical field at voltage $U$ = 9.14 V in the $(x, y)$ plane. (b) The corresponding tilt angle $\varphi$ as a function of time at $U$ = 9.14 V.}
\label{Fig6} 
\end{center}
\end{figure}

To gain further insights into the mechanism favoring vertical orientation of the microbullet, we dispersed particles in 5CB between two ITO slides with planar anchoring, and applied an electric field to dynamically tune the orientation of the 5CB director within the cell. Recall that, in the case of planar cells, microbullets align their long axis parallel to the director. By applying a sinusoidal voltage (in the range of $U$ = 0 - 10 V) at 1 kHz across the ITO electrodes, we created an electric field $\textbf{E}$ perpendicular to the cell walls. We observed a change in the director pattern within the cell under a polarizing microscope as a function of the voltage. The threshold voltage at which molecules began to reorient with the electric field, the Fr\' eedericksz transition was $U \approx$ 4 V. Correspondingly, microbullets rotated to orient perpendicular to the cell walls.  The amplitude of the driving voltage at which this rotation occurred was typically around $U$ $\approx$ 9.14 V (Figure \ref{Fig6} (a)). This latter behavior is reminiscent of the rotation of colloidal platelets dispersed in NLC driven by an external field \cite{Lapointe2010}. Using video microscopy and image analysis we probed the dynamics of the particle rotation by tracking the long axis projection of the bullet in the $(x,y)$ plane. The tilt angle $\varphi$  as a function of time can be deduced directly since we know the particle length (see Figure \ref{Fig6} (b)). When the electric field was switched off, the director of the NLC relaxed to the planar case and the microbullets rotated back to their initial configuration. 
To confirm that a dielectric torque arising from the anisotropy of the dielectric constant due to the shape of colloids \cite{Lapointe2010} was not responsible for the rotation of microbullets, we heated the nematic phase to the isotropic phase and then applied an electric field. In this case, particles did not rotate in response to the electric field, suggesting that microbullet orientation is depended solely on the elastic torque.

When an electric field is used to reorient the nematic director, there is an excess free energy created due to surface anchoring of the particle before it rotates in response. To minimize the excess free energy induced by distortion and anchoring energies, the nematic drives rotation of the bullets from an initial orientation $\varphi$ = $\varphi _0$ to an equilibrium configuration $\varphi$ = $\pi$/2 via an elastic torque \cite{Lapointe2010, Lapointe2004, Lapointe2005, Lapointe2008}. Intriguingly, the rotation can be fit to a growing exponential (solid line in Figure \ref{Fig6} (b)). In contrast, a decaying exponential is predicted near $\varphi$ = $\pi$/2 for a particle rotating in creeping flow owing to a nematic elastic free energy that is harmonic in the orientation angle.


\begin{equation} 
\varphi = \varphi_1 exp \left[ \left(\frac{KLR}{\sigma \eta \zeta}\right)t \right] + \varphi_0 =  \varphi_1 exp \left(\frac{t}{\tau}\right) + \varphi_0              
\end{equation}
with $\varphi_1 \approx$ 5.37 $\times$ $10^{-2}$ rad, $\varphi_0 \approx$ 6.23 $\times$ $10^{-2}$  rad and $\tau \approx$ 0.76 s (the time over which the bullet rotates). 

\begin{figure}
\begin{center}
\includegraphics[width=1\textwidth]{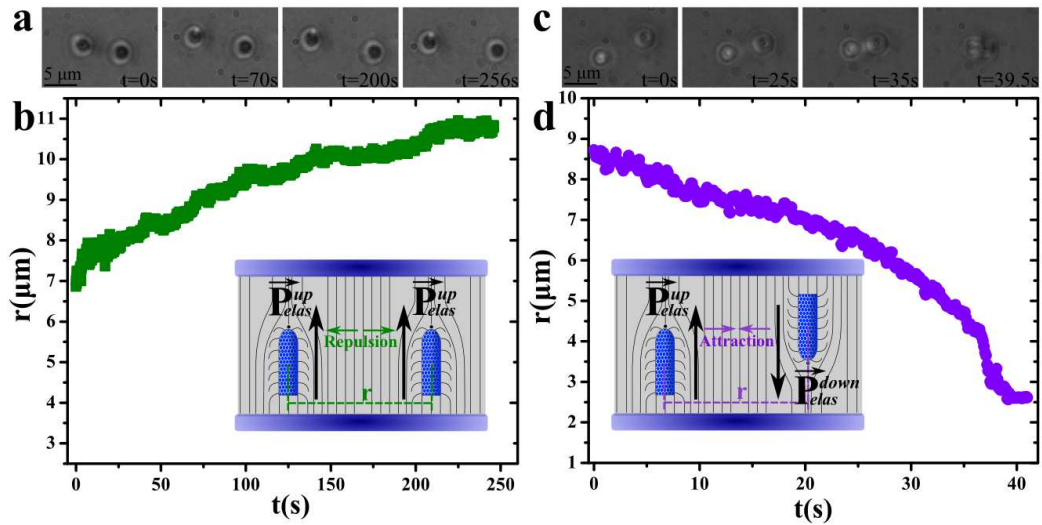}
\caption{(a) Sequential optical images capturing the repulsion between parallel dipoles in a homeotropic NLC cell. (b) Time dependent separation distance, $r$, between two repulsive microbullets when elastic dipoles are parallel. (c) Sequential optical images capturing the attraction between anti-parallel dipoles. (d) Time dependence of the separation distance between two attractive microbullets when elastic dipoles are anti-parallel. Schematics are provided as insets for the possible nematic textures around the microbullets.}
\label{Fig7} 
\end{center}
\end{figure}

Finally, we demonstrated the possibility of creating a dense array of microbullets in homeotropic cells. Microbullets underwent Brownian motion and diffused through the sample over time. As mentioned previously, the height of each colloid was fixed by the boundary conditions in the cell and the orientation of the elastic dipole associated with the particles. For particles in proximity, two distinct behaviors were observed depending on the orientation of their dipoles: when they were parallel they repelled (Figure \ref{Fig7} (a)-(b)) and when they were anti-parallel, they attracted (Figure \ref{Fig7} (c)-(d)). Interactions were recorded and separation distances are tracked over time as shown in Figure \ref{Fig7}.

\begin{figure}
\begin{center}
\includegraphics[width=0.6\textwidth]{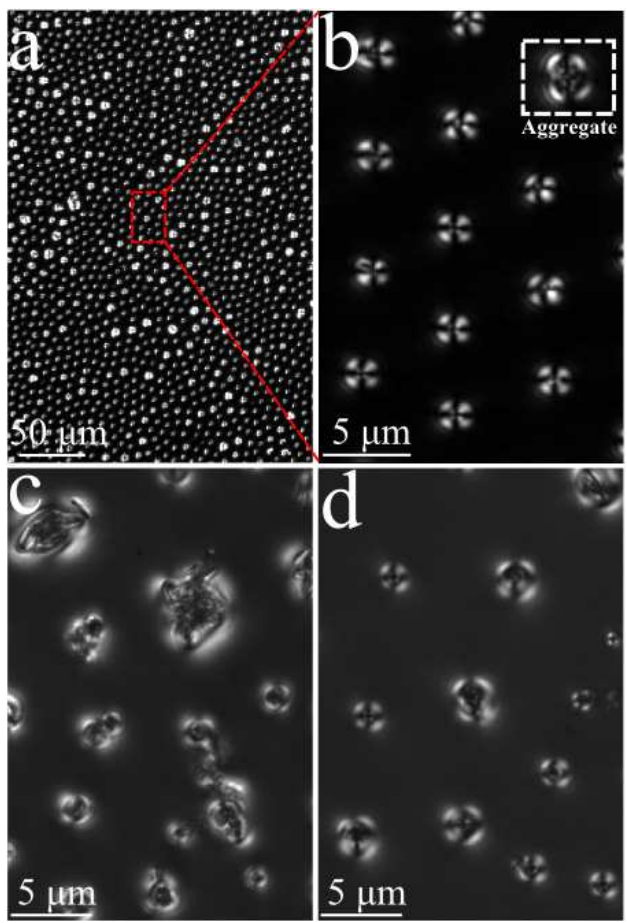}
\caption{(a) Optical microscopy images of patterns formed by a high density of microbullets (3\% wt. in 5CB) dispersed in a homeotropic NLC cell (thickness $h$=25 $\mu$m).(a) Formation of crystal structures with hexagonal symmetry due to the confinement imposed by the cell walls, which coexist with aggregates. (b) A higher magnification of (a). (c) Formation of large aggregates. (d) Formation of a stable ``nematic-like" layer of microbullets: particles are oriented parallel to the director of the LCs but their centers of mass are randomly distributed.}
\label{Fig8} 
\end{center}
\end{figure}

The interactions between microbullets in homeotropic cells leads to a collective  organization of colloids as shown in Figure \ref{Fig8} (a). Large areas of ordered structures with hexagonal symmetry were observed after moderate time scales (overnight) for a density of at least 3 \% wt. in 5CB (Figure \ref{Fig8} (b)). These crystal structures are governed by the elastic repulsive interaction between parallel dipoles similar to the crystal structures observed previously at the air-nematic interface \cite{Gharbi2011}. However, these clusters coexisted with domains of aggregates with complex structures that we attribute to anti-parallel dipoles attracting and consuming their neighbors; the dimers thus formed grew into larger aggregates with time (Figure \ref{Fig8} (c)). Over the course of several days, the separation distance between colloids in the region originally oriented in a hexagonal lattice increased so that particles no longer interacted with their neighbors. A stable ``nematic-like" layer of microbullets is formed (Figure \ref{Fig8} (d)), i.e. all particles are oriented parallel to the director of the LC with centers of mass randomly distributed in the bulk. These results are general, in that they reveal the interactions that occur between cylindrical or rod-like particles with homeotropic anchoring, which also form dipolar defect structures. We note that all patterns disappeared when the system was heated to the isotropic phase, supporting the idea that nematic elasticity is the origin of the particle organization.

\section{Conclusion}
We demonstrated the ability to harness elastic energies to direct anisotropic particle orientation and assembly in NLCs. Under appropriate conditions, spontaneous spatial organization of anisotropic bullet particles occurred. In planar cells, microbullets assembled into 1D chains via dipolar elastic interactions, whereas in homeotropic cells, they rotated to orient their elastic dipoles parallel to the director. By applying an electrical field we showed that anchoring conditions inside the cell can be altered, allowing us to flip microbullets to orient parallel to the director, an effect driven by an elastic torque. In homeotropic cells, depending on the orientation of the dipoles, bullets repelled or attracted each other and formed 2D structures with hexagonal symmetry, a ``nematic-like" phase and aggregates with complex structures, respectively. We believe that the work presented here will offer new insights into the directed assembly of anisotropic particles within complex fluids to develop novel materials and motivate further experimental and theoretical studies to understand the resulting physical properties.

\section*{Acknowledgements}
This work was supported in part by National Science Foundation (NSF) grants CMMI09-00468 and MRSEC DMR11-20901 and a gift from L.J. Bernstein. D.A.B. is supported by an NSF Graduate Research Fellowship. We acknowledge Professor Tom Lubensky (Penn/Physics) for insightful discussion and the Penn Regional Nanotechnology Facility (PRNF) for access to SEM imaging.

\markboth{Taylor \& Francis and I.T. Consultant}{Journal of Modern Optics}

\label{lastpage}

\end{document}